\documentclass[reprint, secnumarabic, amssymb, amsmath, superscriptaddress, aps, prl]{revtex4-2}
\usepackage{graphicx}
\usepackage[colorlinks=true,urlcolor=blue]{hyperref}
\usepackage{xcolor}

\hypersetup{
colorlinks,
linkcolor={red!50!black},
citecolor={blue!50!black},
urlcolor={blue!80!black}
}
\usepackage[normalem]{ulem}
\usepackage{units}

\begin{document}

\title{\textbf{Momentum-resolved EELS study of collective charge excitations in 1$T$-TaS}\textsubscript{\textbf{2}}}

\author{Farzaneh Hoveyda-Marashi}
\affiliation{Department of Physics, University of Illinois, Urbana, Illinois 61801}
\affiliation{Materials Research Laboratory, University of Illinois, Urbana, Illinois 61801}

\author{Xuefei Guo}
\affiliation{Department of Physics, University of Illinois, Urbana, Illinois 61801}
\affiliation{Materials Research Laboratory, University of Illinois, Urbana, Illinois 61801}

\author{Caitlin Kengle}
\affiliation{Department of Physics, University of Illinois, Urbana, Illinois 61801}
\affiliation{Materials Research Laboratory, University of Illinois, Urbana, Illinois 61801}

\author{Camille Bernal-Choban}
\affiliation{Department of Physics, University of Illinois, Urbana, Illinois 61801}
\affiliation{Materials Research Laboratory, University of Illinois, Urbana, Illinois 61801}
\author{Yue Su}
\affiliation{Department of Physics, University of Illinois, Urbana, Illinois 61801}
\affiliation{Materials Research Laboratory, University of Illinois, Urbana, Illinois 61801}

\author{Jin Chen }
\affiliation{Department of Physics, University of Illinois, Urbana, Illinois 61801}
\affiliation{Materials Research Laboratory, University of Illinois, Urbana, Illinois 61801}
\author{Dipanjan Chaudhuri}
\affiliation{Department of Physics, University of Illinois, Urbana, Illinois 61801}
\affiliation{Materials Research Laboratory, University of Illinois, Urbana, Illinois 61801}
\author{Peter Abbamonte}
\affiliation{Department of Physics, University of Illinois, Urbana, Illinois 61801}
\affiliation{Materials Research Laboratory, University of Illinois, Urbana, Illinois 61801}
\email{abbamonte@mrl.illinois.edu}

\begin{abstract}

We use momentum-resolved electron energy-loss spectroscopy (M-EELS) to study the low-energy charge excitations of 1$T$-TaS$_2$ across the nearly commensurate-to–commensurate charge-density-wave (CDW) transition. Single-crystal x-ray diffraction and elastic M-EELS measurements confirm the expected rotation of the CDW wave vector upon entering the commensurate phase. In the nearly commensurate phase, the low-energy M-EELS spectra reveal an acoustic phonon branch and two optical phonon features whose energies and dispersions are broadly consistent with previous calculations and inelastic x-ray measurements. Across the transition, the optical phonon energies remain nearly unchanged, while their spectral intensity develops a pronounced temperature dependence near the CDW ordering wave vector. At higher energies, the finite-momentum charge response undergoes a substantial redistribution of spectral weight below the transition, consistent with the opening of an energy gap. These results demonstrate that M-EELS provides simultaneous access to lattice dynamics and finite-momentum valence band charge excitations in 1$T$-TaS$_2$, revealing their evolution across the commensurate CDW transition.

\end{abstract}

\begin{figure*}
	\center
    \includegraphics[width=1\linewidth]{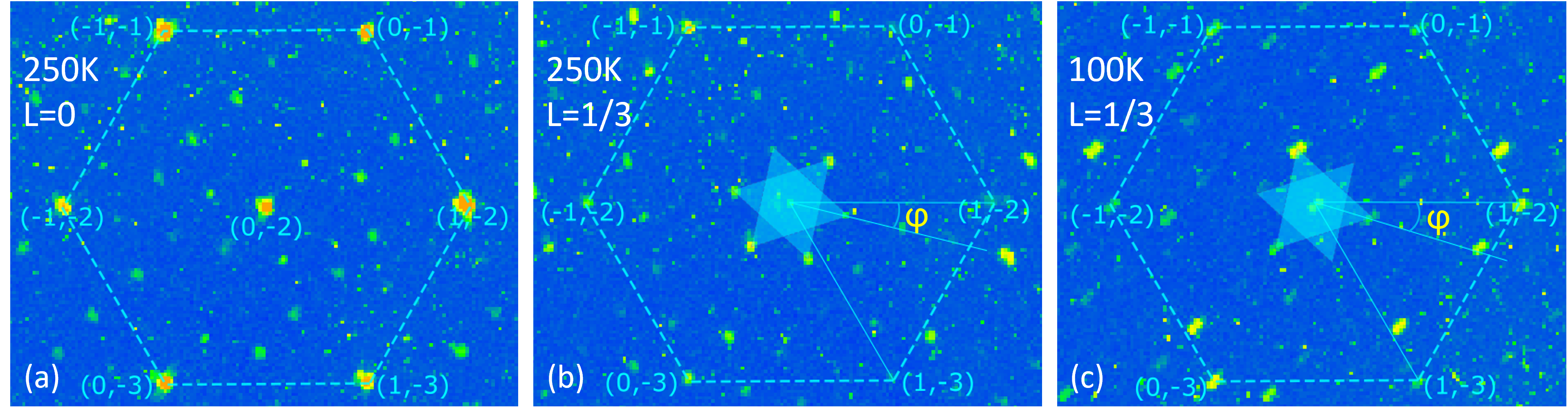}
	\caption{Single-crystal x-ray diffraction measurements of 1$T$-TaS$_2$. (a) Reciprocal-space map at $T=250$ K and $L=$0, showing the structural Bragg reflections. (b,c) Reciprocal-space maps at $L=$1/3 measured at 250 K in the nearly commensurate phase and 100 K in the commensurate phase, respectively. CDW satellite reflections are visible around the structural Bragg peaks. The angle $\varphi$ indicates the orientation of the CDW wave vector, which rotates across the NC–C transition.}
	\label{fig: 1}
\end{figure*}

\maketitle
\section{Introduction}
Charge density waves (CDWs) are a ubiquitous form of broken-symmetry state in correlated electron systems, arising from an intrinsic instability of the electronic structure driven by electron–electron interactions. They have been a subject of sustained interest for decades, most recently because CDW order frequently competes or intertwines with other collective phases, including superconductivity, spin-density-wave order, and electronic nematicity \cite{2009-Berg, 2015-Fradkin}. As a result, CDWs serve as an important model system for understanding how collective electronic degrees of freedom organize in low-dimensional and strongly correlated materials.

Despite decades of study, 
significant questions remain about the collective dynamics of CDWs, largely due to experimental limitations. The relevant excitations occur at meV energy scales and finite momentum \cite{Gruner, 2011-Rossnagel}, a region of phase space traditionally only accessible with inelastic x-ray and neutron scattering. These techniques are mainly sensitive to phonon excitations \cite{2017-Vig,Boothroyd2020}, so behavior of electronic excitations associated with CDWs has been explored very little. 

In recent years, tremendous progress has been made in meV-resolved inelastic electron scattering with high momentum resolution (EELS) \cite{2025-Abbamonte}. In contrast to neutrons and x-rays, EELS is mainly sensitive to electronic excitations, since the scattering from nuclei and core electrons cancel one another \cite{ZhuTafto1996,2017-Vig}. This attribute of EELS has been used to detect Bose condensation of excitons in TiSe$_2$ \cite{Kogar2017} as well as a susceptibility divergence near the CDW transition in ErTe$_3$ \cite{Chaudhuri2025}, opening the possibility of studying electronic properties of CDWs more generally. 

Among CDW materials, the transition metal dichalcogenides (TMDs) have emerged as canonical systems. A prototypical example is TaS\textsubscript{2}, which crystallizes in several polytypes, most notably the 1$T$ and 2$H$ structures. Of these, 1$T$-TaS\textsubscript{2} exhibits an exceptionally rich phase diagram characterized by multiple charge-ordered states \cite{Sipos2008,2011-Rossnagel,1974-Wilson,1975-Scruby, 2017-Law}. Upon cooling, 1$T$-TaS$_2$ undergoes a sequence of CDW transitions. An incommensurate CDW phase first develops below approximately 543 K, followed by a nearly commensurate CDW phase below about 352 K. At still lower temperatures, below 183 K, the system undergoes a first-order transition into a fully commensurate CDW phase characterized by a $\sqrt{13}$ × $\sqrt{13}$ superstructure rotated with respect to the underlying lattice \cite{1971-THOMPSON,Sipos2008,2015-Chen,1977-DISALVO}. In real space, this reconstruction corresponds to the formation of the well-known Star-of-David clusters, in which thirteen Ta atoms distort cooperatively. The transition from the nearly commensurate to the commensurate CDW phase is first order and is widely believed to be accompanied by a Mott metal–insulator transition. In this picture, Coulomb repulsion among the small number of Ta 5d electrons per reconstructed unit cell—electrons that remain partially ungapped in the nearly commensurate phase—drives localization in the commensurate state \cite{1978-Fazekas, 1985-Smith,2006-Rossnagel}. As a result, 1$T$-TaS$_2$ provides a rare and compelling platform in which charge order, electronic correlations, and lattice reconstruction are deeply entangled \cite{1997-Spijkerman}.

Reports of the periodicity of the commensurate phase along the c direction are conflicting. Spijkerman et al. reported an ordered stacking with a periodicity corresponding to $L$=1/3 \cite{1997-Spijkerman}. In contrast, several earlier diffraction studies found no well-defined periodicity along the $c$-axis direction, instead observing diffuse scattering over a broad range of $L$ values, indicative of substantial stacking disorder \cite{1980-Fung,1984-Nakanishi,1984-Tanda}.

Previous studies of collective excitations in  1$T$-TaS$_2$ have focused primarily on the dispersion of acoustic phonon modes. In inelastic neutron scattering measurements, Ziebeck et al. \cite{1977-Ziebeck} observed a pronounced phonon softening in the nearly commensurate phase, consistent with a Kohn anomaly at $q \simeq$0.27 r.l.u. along the $\Gamma$-M direction. Using x-ray thermal diffuse scattering, Machida et al. \cite{2004-Machida} inferred a similar anomaly at $q\simeq$0.28 r.l.u. in the high-temperature incommensurate phase; this feature occurs at a different out-of-plane momentum but corresponds to the same in-plane wave vector. More recently, Liu et al. \cite{2020-Liu} employed inelastic x-ray scattering to directly measure the phonon dispersion and reported an anomaly at $q\simeq$0.3 r.l.u. along $\Gamma$-M, in quantitative agreement with the earlier neutron results in the nearly commensurate phase.

By contrast, considerably less attention has been paid to optical phonons in 1$T$-TaS$_2$. Moreover, while lattice excitations provide important insight into electron-phonon coupling, a complete description of the CDW state also requires measurement of the valence band electronic excitations, such as the opening of a CDW-induced gap. Here, we present a study of 1$T$-TaS$_2$ using momentum-resolved inelastic electron scattering (M-EELS). A key advantage to the EELS technique is that it is intrinsically more sensitive to charge and electronic collective excitations than either inelastic x-ray or neutron scattering \cite{2017-Vig, 2025-Abbamonte,ZhuTafto1996}, providing detailed insight into the behavior of valence band electronic excitations.

\section{Experiment}
Single crystals of 1$T$-TaS$_2$ were commercially obtained from HQGraphene. In order to assure high crystal quality, the crystal structure and charge-density-wave (CDW) order were characterized using single-crystal x-ray diffraction (XRD) with a Bruker X8ApexII (APEX) instrument.
XRD measurements were carried out both above and below the NC–C transition temperature. The results are summarized in Fig. 1, which shows reciprocal-space cuts at different values of the out-of-plane momentum $L$. Above the transition, the characteristic satellite peaks associated with the NC-CDW are clearly observed, confirming the presence of the NC phase. Upon cooling into the commensurate phase, the expected reorientation of the CDW wave vector is observed, providing an unambiguous structural fingerprint of the NC–C transition.

While the underlying crystal exhibits high three-dimensional structural quality, the CDW satellites are observed over a broad range of $L$ values in both phases. This indicates a lack of long-range coherence of the CDW order along the $c$-axis and is consistent with significant stacking faults in the CDW order \cite{1980-Fung,1984-Nakanishi,1984-Tanda}. In one crystal, duplicate reflections, forming an array of twelve satellites around each structural Bragg peak, were observed similar to those reported intermittently in Ref. \cite{2025-Torre}. In this study, we focus on crystals exhibiting the usual six satellite reflections, as shown in Fig. 1, which constituted the majority of the samples.

For M-EELS measurements, 1$T$-TaS$_2$ crystals were mechanically cut into millimeter-scale pieces using a diamond wire saw. Back-reflection Laue diffraction (MWL120) was performed both before and after cutting to verify crystalline quality and to determine the in-plane crystallographic orientation, which was marked on the sample holder. Posts were then glued onto the samples, which were transferred into the UHV system. Samples were cleaved {\it in situ} at a base pressure of $5 \times 10^{-10}$ Torr and transferred directly into the EELS chamber for measurements.

M-EELS experiments were performed using a custom-built, high-resolution instrument based on a reflection-mode HR-EELS spectrometer equipped with a eucentric five-axis goniometer. The eucentric geometry allows independent, automated control of incident and scattered angles while maintaining a fixed sample position, enabling precise and independent control over the in-plane momentum transfer, $q$, and the out-of-plane momentum transfer, $q_z$.
With M-EELS, the momentum transfer $q$ is held constant while scanning the energy loss. Because the electron wave vector satisfies $k \propto \sqrt{E}$, inelastic scattering reduces the magnitude $|k_{\mathrm{out}}|$. To maintain a fixed $q = k_{\mathrm{in}} - k_{\mathrm{out}}$, the scattering geometry is adjusted during the scan by coordinated rotation of $\theta$ and $2\theta$. This procedure is not normally implemented in conventional HR-EELS that lack precise centering adjustments and in which the geometry is fixed during energy scans \cite{2017-Vig,2025-Abbamonte, 2024-Chen}.

The instrument was tuned in the direct-beam configuration to optimize the balance between energy resolution, momentum resolution, and beam current, as described previously \cite{2024-Chen}. Under the tuning conditions used in this work, the energy resolution was $\Delta$E = 5.9 meV as measured in direct beam conditions (without a sample) and 6.2 meV as determined from the full width at half maximum of the elastic line off the TaS$_2$ crystal. The in-plane momentum resolution was $\Delta$q = 0.019 Å\textsuperscript{-1}.
All measurements were performed in reflection geometry, with momentum transfer oriented within the basal plane of the crystal.

Elastic scattering intensity maps, acquired by performing mesh scans of the momentum coordinates $(H,K)$ at zero energy transfer ($\omega$ = 0), are shown in Fig. 2.  In these maps, the CDW modulation is visible as a well-defined peak at finite momentum.
The momentum location of the peak is consistent with the x-ray measurements in the NC phase in Fig. 1. 

The CDW reflection is resolution-limited in the transverse direction, i.e., the direction perpendicular to $q$. In this direction, the momentum width of the peak is comparable to that of the specular reflection measured under identical conditions. This indicates that the observed transverse width is dominated by instrumental resolution rather than intrinsic CDW disorder.

Upon cooling into the commensurate (C) phase, the azimuthal angle of the CDW peak shifts from $\phi$=12\textsuperscript{o} to 15\textsuperscript{o}. This shift is close to x-ray diffraction measurements on the same sample, which show a shift from 13\textsuperscript{o} to 16\textsuperscript{o} (Fig. 1), and agrees with prior reports in the literature within experimental uncertainty. Further, when cooling into the C phase, the longitudinal momentum width of the CDW reflection was found to sharpen (Fig. 2). 

\begin{figure}
	\center
	\includegraphics[width=1\linewidth]{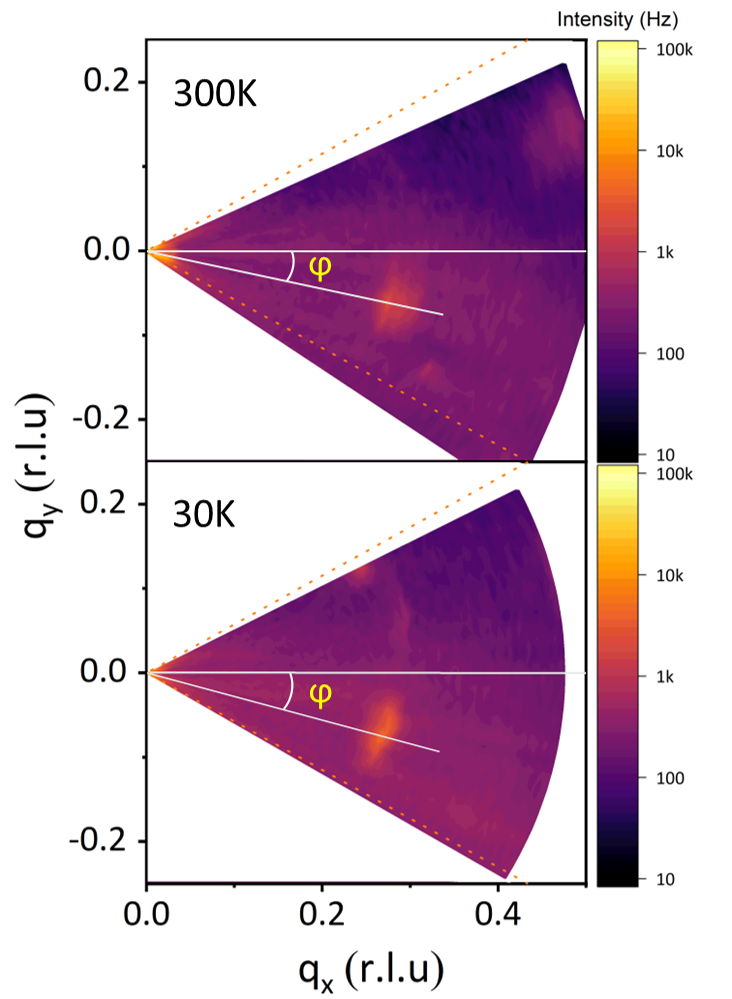}
	\caption{M-EELS momentum maps of elastically scattered electrons acquired at 300 K (top) and 30 K (bottom). Upon cooling from 300 K to 30 K, the CDW wave vector rotates by approximately $3^\circ$. The dashed orange lines indicate the Brillouin-zone boundaries.}
	\label{fig: 2}
\end{figure}
\begin{figure*}[t]
	\center
	\includegraphics[width=1\linewidth]{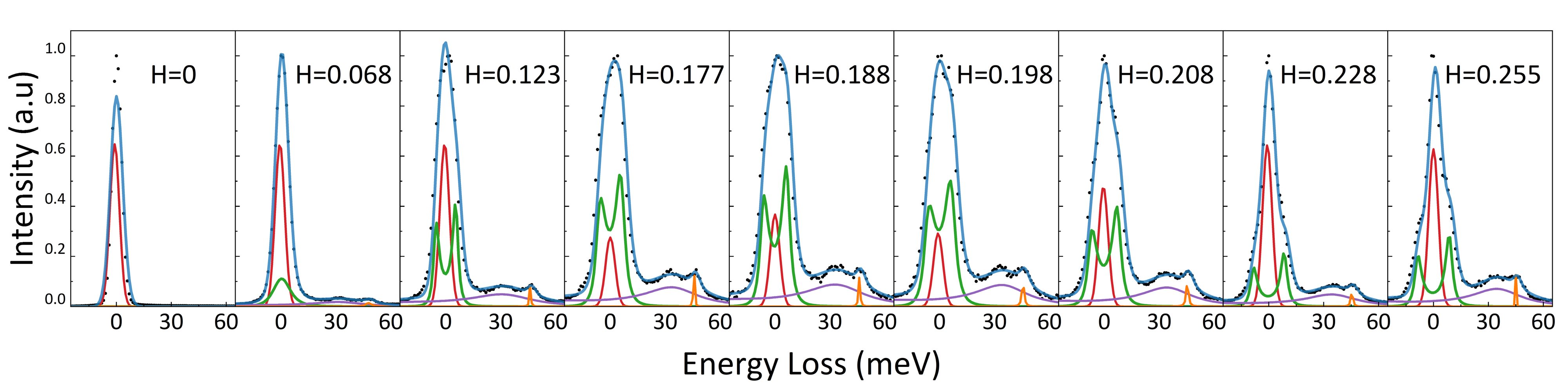}
	\caption{Low-energy M-EELS spectra in the nearly commensurate phase ($T = 300$ K). Fixed-momentum energy-loss scans measured along the CDW direction in reciprocal space. The in-plane momentum transfer is expressed in reciprocal lattice units $(H, K)$, and the scans follow a trajectory parallel to the CDW wave vector, such that both $H$ and $K$ vary simultaneously. The colored spectral components represent the elastic peak in red, the acoustic phonon mode in green, and optical phonon modes in purple and orange. The blue curve shows the overall fit.}
	\label{fig: 3}
\end{figure*} 

Having established the presence and orientation of the CDW order, we proceed to measure the low-energy collective excitations with an energy resolution of $\Delta$E = 5.9 meV.
We begin by examining the low-energy region ($-25 < E < 65$ meV), in which phonon excitations should be visible. Fixed-momentum energy-loss spectra were acquired along the CDW direction ($\phi = 12^\circ$) in the nearly commensurate phase at T = 300 K.

The results are shown in Fig. 3. 
Two optical phonons are visible which, at $H\approx0.07$ r.l.u., have  energies of approximately 28~meV and 48~meV (these features are not visible at $H=0$ because of interference from the elastic line). In addition, an acoustic phonon branch emerges from the elastic line and disperses to higher energy with increasing momentum transfer, $q$.

In order to quantify the phonon dispersions, we fit the  EELS spectra at each momentum transfer to a multicomponent model comprising an elastic line, acoustic phonon and two optical phonons. The elastic line was represented by a Gaussian function with a FWHM of 6.2 meV. Following Ref.~\cite{2023-Kengle}, the acoustic phonon was modeled using a general expression based on time-dependent Landau theory for a scalar mode,
\begin{equation}
    f(A, \mu, \tau, E) = \frac{A E}{E^2 + \left(\mu E^2 - \tau^{-1}\right)^2}.
\end{equation}
where $A$ is an amplitude and $\mu$ and $\tau^{-1}$ are inertial and relaxational parameters, respectively. This form captures both propagating and relaxational dynamics and provides a robust description of the acoustic phonon, even when it is not fully resolved from the elastic line. The optical phonons were modeled using the antisymmetrized damped harmonic oscillator (DHO) response function ~\cite{1997-Fak}, 
\begin{equation}
f(A, E_0, \Gamma, E) = A \left[ \frac{\Gamma}{(E - E_0)^2 + \Gamma^2}-\frac{\Gamma}{(E + E_0)^2 + \Gamma^2}\right].
\end{equation}

The resulting fits are shown as solid lines in Fig.~3, where the individual spectral components are represented as different colors. The extracted phonon dispersions at $T = 300$ K are presented in Fig.~4.
In this figure, the dispersions are compared to the density functional theory calculations  of Liu \textit{et al.}~\cite{2020-Liu}, who calculated phonon dispersions of TaS$_2$ using first-principles methods that properly incorporate  van der Waals interactions.
From this comparison we conclude that some of the phonon features observed in the M-EELS spectra likely correspond to multiple excitations that are unresolved in the measurement. Within this interpretation, overall the agreement between the calculation and the measurement is reasonable both in terms of the overall energies and the magnitudes of the dispersions. 

Next, we investigated how the spectra change as the material is cooled through the commensurate transition at $T_{\mathrm{C-CDW}} = 183$ K, which was shown in Ref. \cite{Sipos2008} to be first order. For this purpose, we measured M-EELS spectra at five different temperatures, chronologically $T =$ 300 K, 30 K, 100 K, 170 K, then 220 K. The results are shown in Fig. 5 for five different momentum values, including $H = 0.233$, which is close to the CDW wave vector. At lower temperatures, the intensity of the acoustic phonon fell below the detection limit because of the reduced Bose occupation factor. So in this figure we focus on the behavior of the optical phonons. We find the energies of these modes do not change appreciably across the transition. The most significant effect, observed only near the CDW ordering wave vector $H_{\mathrm{CDW}}$, is an increase in intensity with decreasing temperature. The magnitude of the increase shows some history dependence; the T=300 K data, which was taken out of sequence, does not follow the same trend as the other temperatures. 
This behavior suggests that optical phonons play a role in the first order commensurate transition. We note that this behavior is quite different from the familiar soft mode scenario characterizing second order structural phase transitions \cite{Gruner}.

Finally, we examine the M-EELS spectra over a wider energy range, -20 meV $< \omega <$ 1000 meV to gain a view of the behavior of electronic excitations (Fig. 6). We focus on the behavior near the CDW ordering wave vector,  $H = 0.233$ for temperatures spanning the commensurate transition. At $T = 220$ K, above the transition, the spectra exhibit a continuum of excitations that decay monotonically with increasing energy. Below $T_{\mathrm{C-CDW}} = 183$ K, a substantial rearrangement of spectral weight is observed, corresponding to the opening of a gap-like feature with magnitude $2\Delta \sim 300$ meV. This value is consistent with previous estimates of the CDW gap from infrared optics, scanning tunneling microscopy, and angle-resolved photoemission spectroscopy \cite{2011-Rossnagel,1985-Smith,1991-Enomoto,2001-Pillo,2002-Gasparov,2004-Aiura,Manzke_1989}. We interpret this feature as the opening of a Mott gap associated with the localization of electrons within the star-of-David clusters in the commensurate phase.

\section{Conclusion}
In summary, we have used momentum-resolved electron energy-loss spectroscopy to study low-energy collective charge excitations in 1$T$-TaS$_2$ across the nearly commensurate to commensurate CDW transition. X-ray diffraction and elastic M-EELS measurements confirm the expected rotation of the CDW wave vector on cooling, establishing that the cleaved surface probed by M-EELS reflects the bulk CDW transition. In the nearly commensurate phase, the low-energy spectra reveal an acoustic branch together with two optical phonon features whose energies and dispersions are broadly consistent with prior first-principles calculations and inelastic x-ray measurements. Across the commensurate transition, the optical phonon energies remain nearly unchanged, but their spectral weight increases strongly near the CDW wave vector, indicating substantial coupling between these modes and the reconstructed CDW state. At higher energies, the finite-momentum charge response undergoes a pronounced redistribution of spectral weight below the transition temperature, producing a gap-like suppression on the scale of a few hundred meV, consistent with previous reports on the commensurate phase. Together, these results show that M-EELS provides a direct probe of the coupled lattice and charge response of 1$T$-TaS$_2$, revealing CDW-coupled phonon excitations at low energy and a gap-like redistribution of spectral weight across the commensurate transition.

\begin{figure}
	\center
	\includegraphics[width=1\linewidth]{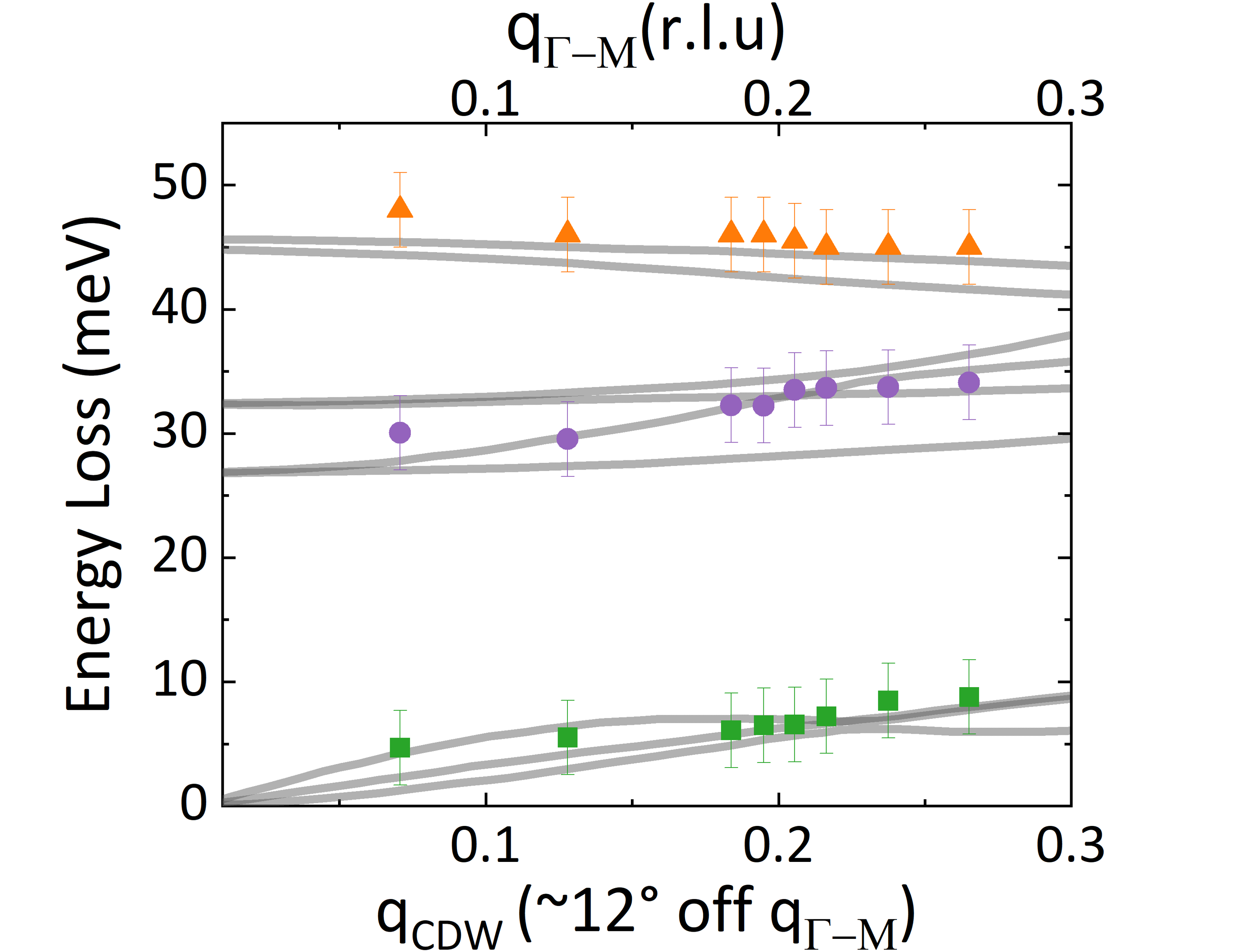}
	\caption{Phonon dispersion of 1$T$-TaS$_2$ in the nearly commensurate phase at $T=300$ K. Experimental mode energies extracted from the M-EELS spectra in Fig. 3 are shown for the acoustic phonon (green squares) and two optical phonon features (purple circles and orange triangles). The measurements were performed along the CDW wave-vector direction, approximately $12^\circ$ from $\Gamma$-M, and are compared with the calculated phonon dispersions of Liu et al. \cite{2020-Liu} along $\Gamma$-M (gray lines). The bottom axis denotes momentum along the experimental CDW direction, while the top axis denotes momentum along $\Gamma$-M used for the calculation.}
	\label{fig: 4}
\end{figure}

\begin{figure}
	\raggedright
\includegraphics[
    width=1.2\linewidth,
    height=0.6\textheight,
    keepaspectratio
]{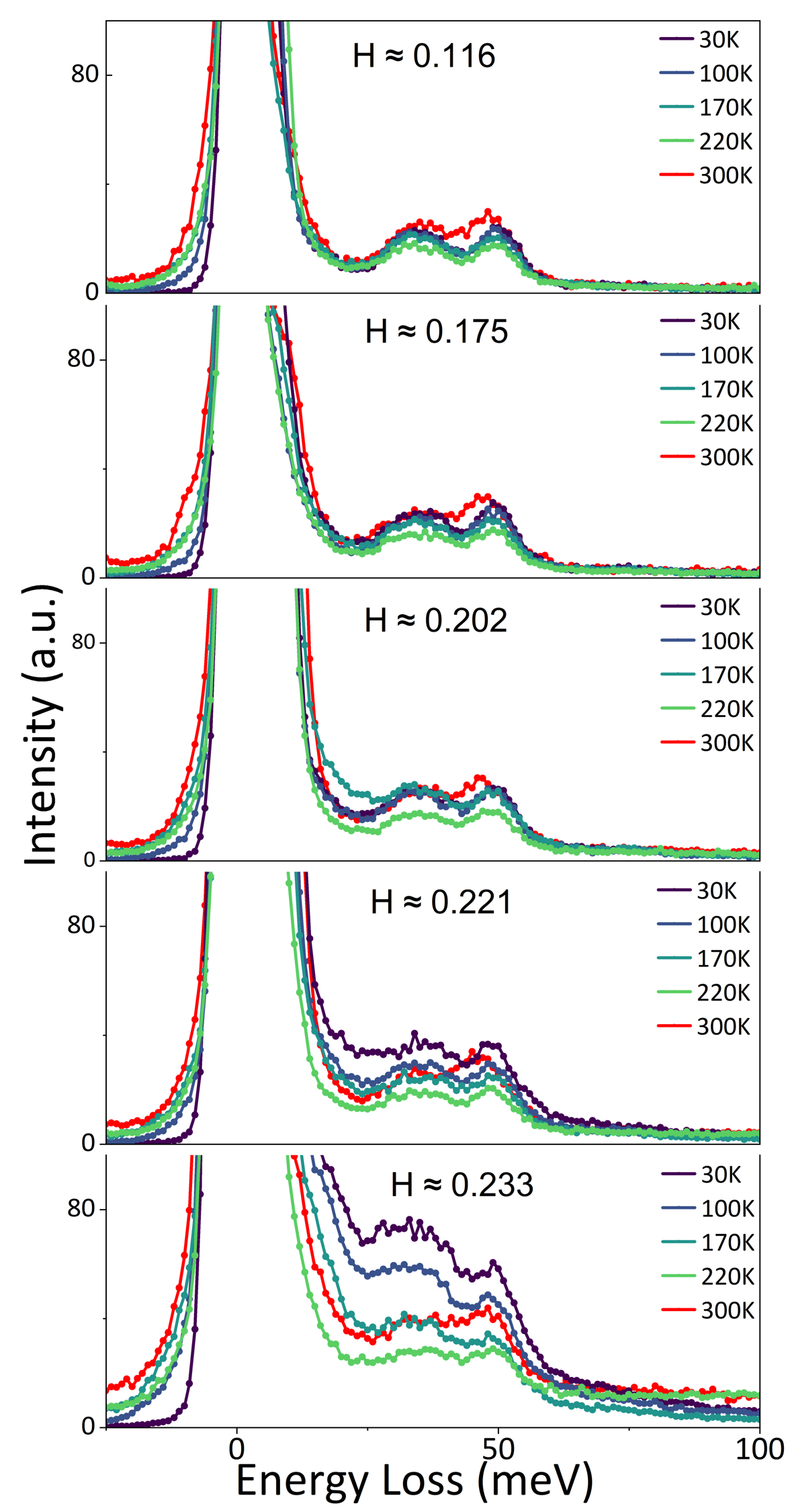}
	\caption{Temperature-dependent M-EELS spectra showing the optical phonon response at selected momenta along the CDW wave-vector direction. Spectra were acquired at $T$=30, 100, 170, 220, and 300 K, spanning the nearly commensurate–commensurate CDW transition. The optical phonon energies remain approximately unchanged across the transition, while their spectral intensity exhibits a pronounced temperature dependence near the CDW ordering wave vector, $H_{\mathrm{CDW}}\approx0.233$ r.l.u.}
	\label{fig: 5}
\end{figure}

\begin{figure}
	\center
	\includegraphics[width=1\linewidth]{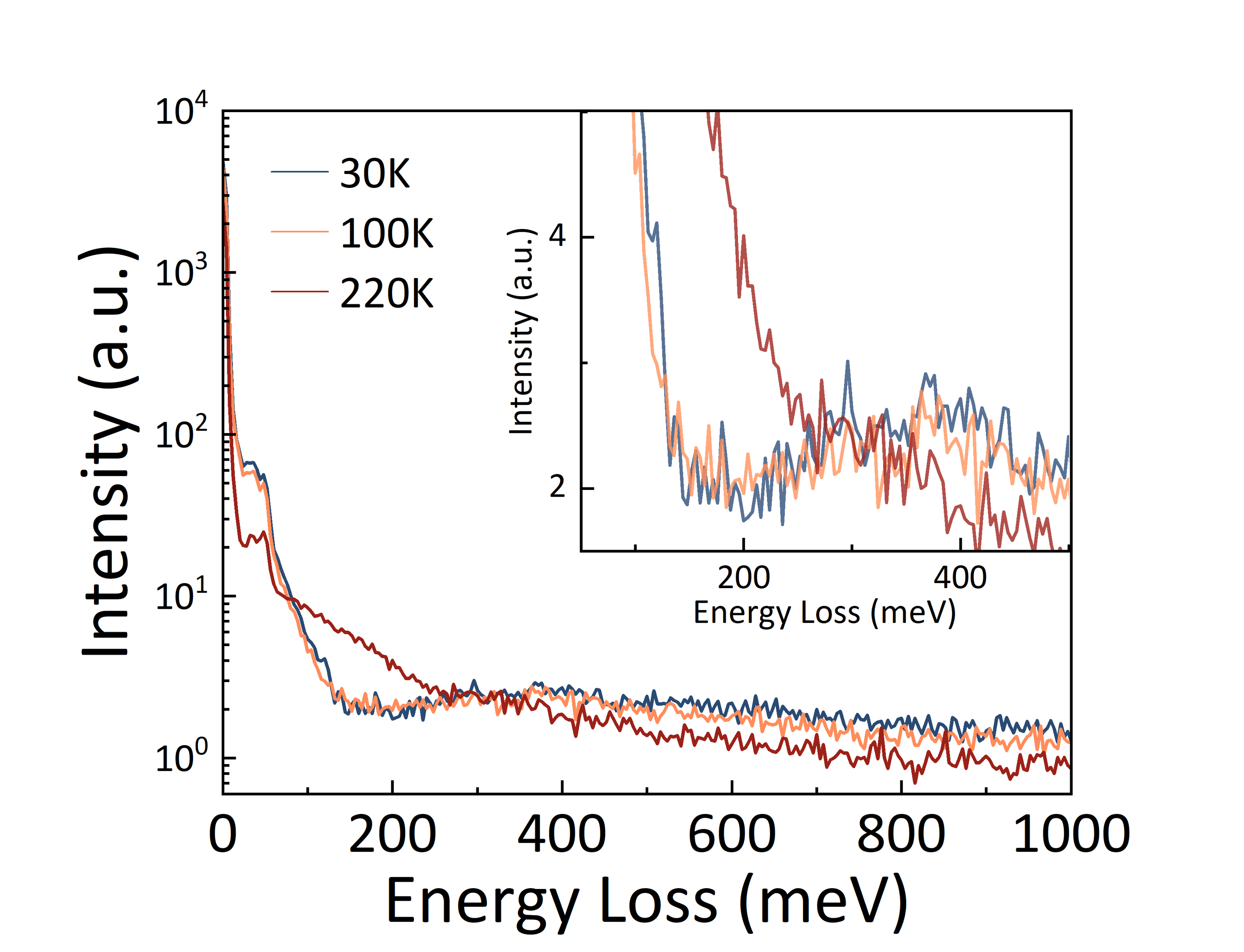}
	\caption{M-EELS energy-loss spectra measured near the CDW ordering wave vector, $H=0.233$ r.l.u., at 30, 100, and 220 K. Below the NC–C transition, the electronic response exhibits a pronounced suppression of spectral weight at low energy, producing a gap-like feature on the scale of a few hundred meV.}
	\label{fig: 6}
\end{figure}

\section{Acknowledgments}

This work was supported by the EPiQS program of the Gordon and Betty Moore Foundation, grant GBMF9452. 

\bibliographystyle{apsrev4-2}
\bibliography{References}

@book{Gruner,
  title = {Density waves in solids},
  author = {Gruner, 
George},
  year = {1994},
  publisher = {CRC Press},
  location = {Boca Raton},
  isbn = {9780429501012},
  pagetotal = {288},
}

@article{2009-Berg,
  title = {Theory of the striped superconductor},
  author = {Berg, Erez and Fradkin, Eduardo and Kivelson, Steven A.},
  journal = {Phys. Rev. B},
  volume = {79},
  issue = {6},
  pages = {064515},
  numpages = {15},
  year = {2009},
  month = {Feb},
  publisher = {American Physical Society},
  doi = {10.1103/PhysRevB.79.064515},
  url = {https://link.aps.org/doi/10.1103/PhysRevB.79.064515}
}

@article{2015-Fradkin,
  title = {Colloquium: Theory of intertwined orders in high temperature superconductors},
  author = {Fradkin, Eduardo and Kivelson, Steven A. and Tranquada, John M.},
  journal = {Rev. Mod. Phys.},
  volume = {87},
  issue = {2},
  pages = {457--482},
  numpages = {26},
  year = {2015},
  month = {May},
  publisher = {American Physical Society},
  doi = {10.1103/RevModPhys.87.457},
  url = {https://link.aps.org/doi/10.1103/RevModPhys.87.457}
}

@article{1977-Ziebeck,
doi = {10.1088/0305-4608/7/7/015},
url = {https://doi.org/10.1088/0305-4608/7/7/015},
year = {1977},
month = {jul},
publisher = {},
volume = {7},
number = {7},
pages = {1139},
author = {K R A Ziebeck and B Dorner and W G Stirling and R Schollhorn},
title = {Kohn anomaly in the 1T2 phase of TaS2},
journal = {Journal of Physics F: Metal Physics}
}

@article{2020-Liu,
author = {Liu, Huili and Yang, Chao and Wei, Bin and Jin, Lei and Alatas, Ahmet and Said, Ayman and Tongay, Sefaattin and Yang, Fan and Javey, Ali and Hong, Jiawang and Wu, Junqiao},
title = {Anomalously Suppressed Thermal Conduction by Electron-Phonon Coupling in Charge-Density-Wave Tantalum Disulfide},
journal = {Advanced Science},
volume = {7},
number = {11},
pages = {1902071},
doi = {https://doi.org/10.1002/advs.201902071},
url = {https://advanced.onlinelibrary.wiley.com/doi/abs/10.1002/advs.201902071},
year = {2020}
}

@article{1997-Spijkerman,
  title = {X-ray crystal-structure refinement of the nearly commensurate phase of $1T\ensuremath{-}{\mathrm{TaS}}_{2}$ in $(3+2)$-dimensional superspace},
  author = {Spijkerman, Albert and de Boer, Jan L. and Meetsma, Auke and Wiegers, Gerrit A. and van Smaalen, Sander},
  journal = {Phys. Rev. B},
  volume = {56},
  issue = {21},
  pages = {13757--13767},
  numpages = {0},
  year = {1997},
  month = {Dec},
  publisher = {American Physical Society},
  doi = {10.1103/PhysRevB.56.13757},
  url = {https://link.aps.org/doi/10.1103/PhysRevB.56.13757}
}

@article{2011-Rossnagel,
doi = {10.1088/0953-8984/23/21/213001},
url = {https://doi.org/10.1088/0953-8984/23/21/213001},
year = {2011},
month = {may},
publisher = {},
volume = {23},
number = {21},
pages = {213001},
author = {Rossnagel, K},
title = {On the origin of charge-density waves in select layered transition-metal dichalcogenides},
journal = {Journal of Physics: Condensed Matter}
}

@article{1975-Scruby,
author = {C. B. Scruby and P. M. Williams and G. S. Parry},
title = {The role of charge density waves in structural transformations of 1T TaS2 },
journal = {The Philosophical Magazine: A Journal of Theoretical Experimental and Applied Physics},
volume = {31},
number = {2},
pages = {255--274},
year = {1975},
publisher = {Taylor \& Francis},
doi = {10.1080/14786437508228930},
URL = { https://doi.org/10.1080/14786437508228930},
}

@article{1984-Nakanishi,
author = {Nakanishi ,Kazuo and Shiba ,Hiroyuki},
title = {Theory of Three-Dimensional Orderings of Charge-Density Waves in 1T-TaX2 (X: S, Se)},
journal = {Journal of the Physical Society of Japan},
volume = {53},
number = {3},
pages = {1103-1113},
year = {1984},
doi = {10.1143/JPSJ.53.1103},
URL = {https://doi.org/10.1143/JPSJ.53.1103}
}

@article{1984-Tanda,
author = {Tanda ,Satoshi and Sambongi ,Takashi and Tani ,Toshiro and Tanaka ,Shoji},
title = {X-Ray Study of Charge Density Wave Structure in 1T-TaS2},
journal = {Journal of the Physical Society of Japan},
volume = {53},
number = {2},
pages = {476-479},
year = {1984},
doi = {10.1143/JPSJ.53.476},
URL = {https://doi.org/10.1143/JPSJ.53.476}
}

@article{1980-Fung,
title = {Application of convergent beam electron diffraction to study the stacking of layers in transition-metal dichalcogenides},
journal = {Physica B+C},
volume = {99},
number = {1},
pages = {47-50},
year = {1980},
issn = {0378-4363},
doi = {https://doi.org/10.1016/0378-4363(80)90208-9},
url = {https://www.sciencedirect.com/science/article/pii/0378436380902089},
author = {K.K. Fung and J.W. Steeds and J.A. Eades}
}

@article{2004-Machida,
author = {Machida ,Yo and Hanashima ,Takayasu and Ohkubo ,Koichi and Yamawaki ,Kouji and Tanaka ,Masahiko and Sasaki ,Satoshi},
title = {Observation of Soft Phonon Modes in 1T-TaS2 by means of X-ray Thermal Diffuse Scattering},
journal = {Journal of the Physical Society of Japan},
volume = {73},
number = {11},
pages = {3064-3069},
year = {2004},
doi = {10.1143/JPSJ.73.3064},
URL = { https://doi.org/10.1143/JPSJ.73.3064}
}

@article{2025-Torre,
author={de la Torre, Alberto
and Wang, Qiaochu
and Masoumi, Yasamin
and Campbell, Benjamin
and Riffle, Jake V.
and Balasundaram, Dushyanthini
and Vora, Patrick M.
and Ruff, Jacob P. C.
and Fiete, Gregory A.
and Hollen, Shawna M.
and Plumb, Kemp W.},
title={Dynamic phase transition in 1T-TaS2 via a thermal quench},
journal={Nature Physics},
year={2025},
month={Aug},
day={01},
volume={21},
number={8},
pages={1267-1274},
issn={1745-2481},
doi={10.1038/s41567-025-02938-1},
url={https://doi.org/10.1038/s41567-025-02938-1}
}

@article{1985-Smith,
doi = {10.1088/0022-3719/18/16/013},
url = {https://doi.org/10.1088/0022-3719/18/16/013},
year = {1985},
month = {jun},
publisher = {},
volume = {18},
number = {16},
pages = {3175},
author = {N V Smith and S D Kevan and F J DiSalvo},
title = {Band structures of the layer compounds 1T-TaS2 and 2H-TaSe2 in the presence of commensurate charge-density waves},
journal = {Journal of Physics C: Solid State Physics}
}

@article{1991-Enomoto,
    author = {Enomoto, H. and Ozaki, H. and Suzuki, M. and Fujii, T. and Yamaguchi, M.},
    title = {Angle‐resolved tunneling spectroscopy of the charge density wave density of states in 1T‐TaS2},
    journal = {Journal of Vacuum Science I\& Technology B: Microelectronics and Nanometer Structures Processing, Measurement, and Phenomena},
    volume = {9},
    number = {2},
    pages = {1022-1026},
    year = {1991},
    month = {03},
    issn = {1071-1023},
    doi = {10.1116/1.585251},
    url = {https://doi.org/10.1116/1.585251},
}

@article{2001-Pillo,
  title = {Fine structure in high-resolution photoemission spectra of quasi-two-dimensional $1T\ensuremath{-}{\mathrm{TaS}}_{2}$},
  author = {Pillo, Th. and Hayoz, J. and Naumovi\ifmmode \acute{c}\else \'{c}\fi{}, D. and Berger, H. and Perfetti, L. and Gavioli, L. and Taleb-Ibrahimi, A. and Schlapbach, L. and Aebi, P.},
  journal = {Phys. Rev. B},
  volume = {64},
  issue = {24},
  pages = {245105},
  numpages = {4},
  year = {2001},
  month = {Dec},
  publisher = {American Physical Society},
  doi = {10.1103/PhysRevB.64.245105},
  url = {https://link.aps.org/doi/10.1103/PhysRevB.64.245105}
}

@article{2002-Gasparov,
  title = {Phonon anomaly at the charge ordering transition in $1T\ensuremath{-}{\mathrm{TaS}}_{2}$},
  author = {Gasparov, L. V. and Brown, K. G. and Wint, A. C. and Tanner, D. B. and Berger, H. and Margaritondo, G. and Ga\'al, R. and Forr\'o, L.},
  journal = {Phys. Rev. B},
  volume = {66},
  issue = {9},
  pages = {094301},
  numpages = {5},
  year = {2002},
  month = {Sep},
  publisher = {American Physical Society},
  doi = {10.1103/PhysRevB.66.094301},
  url = {https://link.aps.org/doi/10.1103/PhysRevB.66.094301}
}

@article{2004-Aiura,
  title = {Increase in charge-density-wave potential of $1T\text{\ensuremath{-}}{\mathrm{TaS}}_{x}{\mathrm{Se}}_{2\ensuremath{-}x}$},
  author = {Aiura, Y. and Hase, I. and Yagi-Watanabe, K. and Bando, H. and Ozawa, K. and Tanaka, K. and Kitagawa, R. and Maruyama, S. and Iwase, T. and Nishihara, Y. and Horiba, K. and Shiino, O. and Oshima, M. and Nakatake, M. and Kubota, M. and Ono, K.},
  journal = {Phys. Rev. B},
  volume = {69},
  issue = {24},
  pages = {245123},
  numpages = {7},
  year = {2004},
  month = {Jun},
  publisher = {American Physical Society},
  doi = {10.1103/PhysRevB.69.245123},
  url = {https://link.aps.org/doi/10.1103/PhysRevB.69.245123}
}

@article{1974-Wilson,
  title = {Charge-Density Waves in Metallic, Layered, Transition-Metal Dichalcogenides},
  author = {Wilson, J. A. and Di Salvo, F. J. and Mahajan, S.},
  journal = {Phys. Rev. Lett.},
  volume = {32},
  issue = {16},
  pages = {882--885},
  numpages = {0},
  year = {1974},
  month = {Apr},
  publisher = {American Physical Society},
  doi = {10.1103/PhysRevLett.32.882},
  url = {https://link.aps.org/doi/10.1103/PhysRevLett.32.882}
}

@Article{2017-Vig,
	title={{Measurement of the dynamic charge response of materials using low-energy, momentum-resolved electron energy-loss spectroscopy (M-EELS)}},
	author={Sean Vig and Anshul Kogar and Matteo Mitrano and Ali A. Husain and Vivek Mishra and Melinda S. Rak and Luc Venema and Peter D. Johnson and Genda D. Gu and Eduardo Fradkin and Michael R. Norman and Peter Abbamonte},
	journal={SciPost Phys.},
	volume={3},
	pages={026},
	year={2017},
	publisher={SciPost},
	doi={10.21468/SciPostPhys.3.4.026},
	url={https://scipost.org/10.21468/SciPostPhys.3.4.026},
}

@article{2025-Abbamonte,
   author = "Abbamonte, Peter and Fink, Jörg",
   title = "Collective Charge Excitations Studied by Electron Energy-Loss Spectroscopy", 
   journal= "Annual Review of Condensed Matter Physics",
   year = "2025",
   volume = "16",
   number = "Volume 16, 2025",
   pages = "465-480",
   doi = "https://doi.org/10.1146/annurev-conmatphys-032822-044125",
   url = "https://www.annualreviews.org/content/journals/10.1146/annurev-conmatphys-032822-044125",
   publisher = "Annual Reviews",
   issn = "1947-5462",
   type = "Journal Article",
  }

@article{2024-Chen,
  title = {Consistency between reflection momentum-resolved electron energy loss spectroscopy and optical spectroscopy measurements of the long-wavelength density response of ${\mathrm{Bi}}_{2}{\mathrm{Sr}}_{2}{\mathrm{CaCu}}_{2}{\mathrm{O}}_{8+x}$},
  author = {Chen, Jin and Guo, Xuefei and Boyd, Christian and Bettler, Simon and Kengle, Caitlin and Chaudhuri, Dipanjan and Hoveyda, Farzaneh and Husain, Ali and Schneeloch, John and Gu, Genda and Phillips, Philip and Uchoa, Bruno and Chiang, Tai-Chang and Abbamonte, Peter},
  journal = {Phys. Rev. B},
  volume = {109},
  issue = {4},
  pages = {045108},
  numpages = {9},
  year = {2024},
  month = {Jan},
  publisher = {American Physical Society},
  doi = {10.1103/PhysRevB.109.045108},
  url = {https://link.aps.org/doi/10.1103/PhysRevB.109.045108}
}

@article{2023-Kengle,
  title = {Non-RPA behavior of the valence plasmon in ${\mathrm{SrTi}}_{1\ensuremath{-}x}{\mathrm{Nb}}_{x}{\mathrm{O}}_{3}$},
  author = {Kengle, Caitlin S. and Rubeck, Samantha I. and Rak, Melinda and Chen, Jin and Hoveyda, Faren and Bettler, Simon and Husain, Ali and Mitrano, Matteo and Edelman, Alexander and Littlewood, Peter and Chiang, Tai-Chang and Mahmood, Fahad and Abbamonte, Peter},
  journal = {Phys. Rev. B},
  volume = {108},
  issue = {20},
  pages = {205102},
  numpages = {9},
  year = {2023},
  month = {Nov},
  publisher = {American Physical Society},
  doi = {10.1103/PhysRevB.108.205102},
  url = {https://link.aps.org/doi/10.1103/PhysRevB.108.205102}
}

@article{1997-Fak,
title = {Phonon line shapes and excitation energies},
journal = {Physica B: Condensed Matter},
volume = {234-236},
pages = {1107-1108},
year = {1997},
note = {Proceedings of the First European Conference on Neutron Scattering},
issn = {0921-4526},
doi = {https://doi.org/10.1016/S0921-4526(97)00121-X},
url = {https://www.sciencedirect.com/science/article/pii/S092145269700121X},
author = {B. Fåk and B. Dorner}
}

@article{2017-Law,
author = {K. T. Law  and Patrick A. Lee },
title = {1T-TaS<sub>2</sub> as a quantum spin liquid},
journal = {Proceedings of the National Academy of Sciences},
volume = {114},
number = {27},
pages = {6996-7000},
year = {2017},
doi = {10.1073/pnas.1706769114},
URL = {https://www.pnas.org/doi/abs/10.1073/pnas.1706769114}}

@article{2015-Chen,
  title = {Influence of Ti doping on the incommensurate charge density wave in $1T\text{\ensuremath{-}}{\mathrm{TaS}}_{2}$},
  author = {Chen, X. M. and Miller, A. J. and Nugroho, C. and de la Pe\~na, G. A. and Joe, Y. I. and Kogar, A. and Brock, J. D. and Geck, J. and MacDougall, G. J. and Cooper, S. L. and Fradkin, E. and Van Harlingen, D. J. and Abbamonte, P.},
  journal = {Phys. Rev. B},
  volume = {91},
  issue = {24},
  pages = {245113},
  numpages = {6},
  year = {2015},
  month = {Jun},
  publisher = {American Physical Society},
  doi = {10.1103/PhysRevB.91.245113},
  url = {https://link.aps.org/doi/10.1103/PhysRevB.91.245113}
}

@article{ZhuTafto1996,
  title = {Direct Imaging of Charge Modulation},
  author = {Zhu, Yimei and Tafto, J.},
  journal = {Phys. Rev. Lett.},
  volume = {76},
  issue = {3},
  pages = {443--446},
  numpages = {0},
  year = {1996},
  month = {Jan},
  publisher = {American Physical Society},
  doi = {10.1103/PhysRevLett.76.443},
  url = {https://link.aps.org/doi/10.1103/PhysRevLett.76.443}
}

@article{Kogar2017,
    author = {Anshul Kogar  and Melinda S. Rak  and Sean Vig  and Ali A. Husain  and Felix Flicker  and Young Il Joe  and Luc Venema  and Greg J. MacDougall  and Tai C. Chiang  and Eduardo Fradkin  and Jasper van Wezel  and Peter Abbamonte },
    title = {Signatures of exciton condensation in a transition metal dichalcogenide},
    journal = {Science},
    volume = {358},
    number = {6368},
    pages = {1314-1317},
    year = {2017},
doi = {10.1126/science.aam6432},
URL = {https://www.science.org/doi/abs/10.1126/science.aam6432}
}

@article{Chaudhuri2025,
    author = {Dipanjan Chaudhuri  and Qianni Jiang  and Xuefei Guo  and Jin Chen  and Caitlin S. Kengle  and Farzaneh Hoveyda-Marashi  and Camille Bernal-Choban  and Niels de Vries  and Tai-Chang Chiang  and Eduardo Fradkin  and Ian R. Fisher  and Peter Abbamonte },
    title = {Measurement of the dynamic charge susceptibility near the charge density wave transition in ErTe<sub>3</sub>},
    journal = {Proceedings of the National Academy of Sciences},
    volume = {122},
    number = {25},
    pages = {e2424430122},
    year = {2025},
    doi = {10.1073/pnas.2424430122},
    URL = {https://www.pnas.org/doi/abs/10.1073/pnas.2424430122},
}

@book{Boothroyd2020,
  author = {A. T. Boothroyd},
  publisher = {Oxford University Press},
  title = {Principles of Neutron Scattering from Condensed Matter},
  year = {2020}
}

@Article{Sipos2008,
    author={Sipos, B. and Kusmartseva, A. F. and Akrap, A. and Berger, H. and Forr{\'o}, L. and Tuti{\v{s}}, E.},
    title={From Mott state to superconductivity in 1T-TaS2},
    journal={Nature Materials},
    year={2008},
    month={Dec},
    day={01},
    volume={7},
    number={12},
    pages={960-965},
    issn={1476-4660},
    doi={10.1038/nmat2318},
    url={https://doi.org/10.1038/nmat2318}
}

@article{1977-DISALVO,
title = {The low temperature electrical properties of 1T-TaS2},
journal = {Solid State Communications},
volume = {23},
number = {11},
pages = {825-828},
year = {1977},
issn = {0038-1098},
doi = {https://doi.org/10.1016/0038-1098(77)90961-9},
url = {https://www.sciencedirect.com/science/article/pii/0038109877909619},
author = {F.J. {Di Salvo} and J.E. Graebner}
}

@article{1971-THOMPSON,
title = {Transitions between semiconducting and metallic phases in 1-T TaS2},
journal = {Solid State Communications},
volume = {9},
number = {13},
pages = {981-985},
year = {1971},
issn = {0038-1098},
doi = {https://doi.org/10.1016/0038-1098(71)90444-3},
url = {https://www.sciencedirect.com/science/article/pii/0038109871904443},
author = {A.H. Thompson and R.F. Gamble and J.F. Revelli}
}

@article{1978-Fazekas,
author = {P. Fazekas and E. Tosatti},
title = {Electrical, structural and magnetic properties of pure and doped 1T-TaS2 },
journal = {Philosophical Magazine B},
volume = {39},
number = {3},
pages = {229--244},
year = {1979},
publisher = {Taylor \& Francis},
doi = {10.1080/13642817908245359},
URL = {https://doi.org/10.1080/13642817908245359},
}

@article{Manzke_1989,
doi = {10.1209/0295-5075/8/2/015},
url = {https://doi.org/10.1209/0295-5075/8/2/015},
year = {1989},
month = {jan},
publisher = {},
volume = {8},
number = {2},
pages = {195},
author = {R. Manzke and T. Buslaps and B. Pfalzgraf and M. Skibowski and O. Anderson},
title = {On the Phase Transitions in 1T-TaS2},
journal = {Europhysics Letters}
}

@article{2006-Rossnagel,
  title = {Spin-orbit coupling in the band structure of reconstructed $1T\text{\ensuremath{-}}{\mathrm{TaS}}_{2}$},
  author = {Rossnagel, K. and Smith, N. V.},
  journal = {Phys. Rev. B},
  volume = {73},
  issue = {7},
  pages = {073106},
  numpages = {3},
  year = {2006},
  month = {Feb},
  publisher = {American Physical Society},
  doi = {10.1103/PhysRevB.73.073106},
  url = {https://link.aps.org/doi/10.1103/PhysRevB.73.073106}
}

\end{document}